\documentclass[12pt, a4paper]{article}

\usepackage{amsmath}
\usepackage{amsfonts}
\usepackage{amssymb}
\usepackage{graphicx, rotating}
\usepackage{epstopdf}
\usepackage{epsfig}
\usepackage{latexsym}
\usepackage{graphicx}
\usepackage{color}
\usepackage{amsmath,bm,amssymb}
\usepackage{cite}
\usepackage{slashed}
\usepackage{hyperref}
\hypersetup{colorlinks, citecolor=bluscuro, linkcolor=black, urlcolor=bluscuro}
\definecolor{rossos}{cmyk}{0,1,1,0.55}
\definecolor{bluscuro}{rgb}{0.15, 0.2, .85}
\definecolor{bluchiaro}{cmyk}{1,.3,0.,0.1}

\newcommand{\be}{\begin{equation}}
\newcommand{\ee}{\end{equation}}
\newcommand{\bea}{\begin{eqnarray}}
\newcommand{\eea}{\end{eqnarray}}

\def\Tr{{\rm Tr\,}}

\def\CO{\cal O}
\def\CN{\cal N}

\begin{document}

\begin{titlepage}
\begin{flushright}
IFT-UAM/CSIC-19-98

\end{flushright}
\vspace{.3in}

\vspace{1cm}
\begin{center}
{\Large\bf\color{black} On The Evolution Of Operator Complexity Beyond Scrambling}\\

\bigskip\color{black}
\vspace{1cm}{
{\large J.~L.~F. Barb\'on$^a$, E. ~Rabinovici$^{b,c}$,  R.~Shir$^{b}$, R.~Sinha$^{a}$}
\vspace{0.3cm}
} \\[7mm]
{\it {$^a$\, Instituto de Fisica Teorica IFT-UAM/CSIC, Cantoblanco 28049, Madrid. Spain}}\\
{\it {$^b$\, Racah Institute, The Hebrew University. Jerusalem 9190401, Israel}}\\
{\it {$^c$\, Institut des Hautes Etudes Scientifiques, 91440 Bures-sur-Yvette, France}}
\end{center}
\bigskip

\vspace{.4cm}

\begin{abstract}
We study operator complexity on various time scales with emphasis on those much larger than the scrambling period. We use, for systems with a large but finite number of degrees of freedom, the notion of K-complexity employed in \cite{altman}  for 
infinite systems. We present evidence that K-complexity of ETH operators has indeed the  character associated with the bulk time evolution of extremal volumes and  actions. 
Namely, after a period of exponential growth during the scrambling period the K-complexity increases only linearly with time for exponentially long times 
in terms of the entropy, and it eventually saturates at a constant value also exponential in terms of the entropy.  This constant value depends on the Hamiltonian and the operator but not on any extrinsic tolerance parameter.
Thus K-complexity deserves to be an entry in the AdS/CFT dictionary. 
Invoking a concept of K-entropy and some numerical examples we also discuss
the extent to which the long period of linear complexity growth entails an efficient
randomization of operators.

\end{abstract}
\bigskip

\end{titlepage}


\section{Introduction}

\noindent

Quantum complexity has been proposed as a new entry in the holographic dictionary  (see for instance \cite{bsuss, suss3} and references therein).  The underlying idea is to characterize the entanglement of a state in an `optimal' way, with respect to some simple building blocks, such as gates in a quantum circuit model or more generally a tensor network. Complexity can then be defined as the size of the smallest circuit or tensor network which approximates the state, given some prescribed set of gates or fundamental tensors (see for instance \cite{nielsen}). 

The quantum circuit model leads naturally to a notion of  complexity which is extensive in the number of degrees of freedom, $S$, and furthermore grows linearly in time, for a period  much longer than any ordinary thermalization time scale:
\be\label{ratec}
C[\Psi_t] \propto S\, \cdot{ t\over \beta} \;,
\ee
with $\beta$ an effective time step for state-vector orthogonality, i.e.  $\left\langle \Psi_t \,| \Psi_{t+\beta} \right\rangle \approx 0$. 
This linear growth is to be matched to
the linear growth of spacelike volumes inside a black hole of entropy $S$ and inverse Hawking temperature $\beta$ \cite{epr=er, harmal}. 
 
 An important question is whether the so-defined  complexity has an upper bound. In the quantum models with a finite set of qubits, a computation is regarded as finished when the target state is approximated within some {\it a priori} tolerance $\epsilon$, with respect to a standard metric on the space of states. Complexities defined with such an implicit dependence on the tolerance parameter are bounded by the number of $\epsilon$-cells in the space of states, which scales exponentially with the number of qubits: 
 \be\label{bcom}
 C^{(\epsilon)} < \exp\left(c\,S \,\log(1/\epsilon)\right)\;, 
\ee
where $c$  is a numerical constant of $O(1)$, and we neglect various polynomial corrections to this expression \cite{nielsenchuang}. For any linearly rising complexity,  the bound (\ref{bcom}) is then attained over time scales exponential in the entropy $S$,  similar to the Heisenberg time scale
$
t_H \sim \beta \,e^S\;,
$
which controls the randomization of a quantum state under time evolution, except for the occurrence of the factor $\log (1/\epsilon)$.

It is not clear how to interpret the complexity bound on the gravitational side, since the tolerance parameter  $\epsilon$ lacks a concrete physical interpretation.  The extremal volume through a wormhole grows with no limit in an eternal black hole geometry, only disturbed by non-perturbative effects such as tunneling transitions. If these fluctuations affect the complexity in a similar manner as they affect correlation functions, the relevant time scale for complexity saturation should be the Heisenberg time $t_H$,  which has no dependence on the tolerance parameter \cite{eternal, br}. 

The dependence on $\epsilon$ finds its origin on the notion of a metric over the space of states.  It would be interesting to have a definition of complexity which does not rely on some effective volume, but rather depends on an effective dimensionality. A suggestion in this direction is to  switch  from the Schr\"odinger picture to the Heisenberg picture and characterize complexity in terms of the size of an operator with respect to some local basis of the operator algebra. In a system of qubits, we can think of the Pauli operators at each qubit as generators of the operator algebra, and the size of a given operator may be defined as the average number of non-identity factors in this representation. 

Operator growth and the relation to thermalization has been discussed as a criterion for quantum chaos in model systems for fast scramblers, such as the SYK model \cite{syk}, \cite{stanford, qi}. The authors of ref. \cite{altman} have proposed a variant of these notions which uses an operator basis adapted to the time evolution of operators, rather than an {\it a priori} basis of the operator algebra. Starting from some initial operator ${\cal O}_0$, one can envision the Heisenberg flow on the space of operators, ${\cal O}_t = e^{itH} \,{\cal O}_0 \, e^{-itH}$, whose Taylor expansion with respect to the time variable is generated by the set of nested commutators of ${\cal O}_0$ with the Hamiltonian.  Using these nested commutators as linear generators of the operator algebra one can  describe the Heisenberg flow as gradually accesing a growing subspace of the operator space. K-complexity is defined  in \cite{altman} as an effective dimension of this growing subspace. 

Using the SYK model   as a benchmark model for a fast scrambler, it has been shown in \cite{altman} that K-complexity is similar to other definitions of operator size, when working in the thermodynamic limit.  In this paper, we move away from the thermodynamic limit and study the regime of very long times, much larger than the scrambling time, when operator size ceases to be a useful characterization of complexity. We show that K-complexity continues to grow at a linear rate in this post-scrambling period  until it saturates just below the total dimensionality of the operator space, of order $e^{O(S)}$. Because of the linear rate, this saturation occurs on time scales exponential in the number of degrees of freedom, roughly similar to the Heisenberg time. Both the complexity upper bound and the saturation time scale stand without any reference to a coarse-graining parameter. 

This paper is organized as follows. In section 2 we review the basic concepts and notation elements of K-complexity. In section 3 we
discuss the K-complexity of scrambling systems with a finite number of degrees of freedom, with special emphasis on the post-scrambling evolution.  We establish the linear growth of K-complexity and the saturation time scale. In section 4 we define the notion of K-entropy, as a measure of the degree of randomization of the Heisenberg flow, and argue on the basis of some numerical estimates that such randomization is expected to occur in order of magnitude. Section 5 brings the conclusions and a number of open questions suggested by our work.

\section{Review of K-complexity}

\noindent

We begin with a review of K-complexity and a description of the notational conventions to be used in this paper. The main reference for this section is \cite{altman}. 

Given the Hamiltonian  of a lattice  system, $H$, and a  particular initial operator ${\cal O}_0$, one defines a linearly independent set of  operators  ${\cal O}_n$  in terms of the  $n$-times nested commutators $[H, [H, \cdots, [H, {\cal O}_0], \dots ],]$, conveniently improved into an orthonormal set, known as the Krylov basis. This choice is motivated by the time evolution, since the nested commutators determine the time Taylor expansion of the Heisenberg operator ${\cal O}_t = e^{itH} \,{\cal O}_0 \,e^{-itH}$.  The orthonormality  can be defined with respect to any non-degenerate inner product in the operator algebra, such as the trace inner product
\be\label{inner}
({\cal O} | {\cal O}') = \frac{ 1}{ \CN}\; \Tr ( \CO^\dagger \CO')\;, \qquad \CN \equiv \Tr {\bf 1}
\;.
\ee
The construction of the Krylov basis runs iteratively as follows. From the initial operator ${\cal O}_0 = A_0$ we define $A_1 = [H, A_0]$ and ${\cal O}_1 = b_1^{-1} A_1$, with $b_1 = \sqrt{(A_1 | A_1)}$.  From here onwards,  given ${\cal O}_{n-1}$ and ${\cal O}_{n-2}$ we define 
\be\label{itera}
A_n = [H, {\cal O}_{n-1}] -b_{n-1} \,{\cal O}_{n-2}
\;,
\ee
and 
\be\label{norm}
{\cal O}_n = b_n^{-1}\, A_n\;, \qquad b_n = \sqrt{(A_n | A_n )}\;.
\ee
The adjoint action of the Hamiltonian is almost diagonal on the orthonormal set ${\cal O}_n$:
\be\label{hamaj}
[H, {\cal O}_n ] = b_{n+1}  \,{\cal O}_{n+1} + b_n \, {\cal O}_{n-1} \;,
\ee
where the   non-negative matrix elements  $b_n$ are called Lanczos coefficients. It is useful to exploit the notation (\ref{inner}) to introduce a vector space of operators, $|\CO )$,  with dimension of order ${\cal N}^{\,2}$. The adjoint action of the Hamiltonian in (\ref{hamaj}) introduces a linear operator in this space known as the {\it Liouvillian}, defined as
\be\label{liou}
{\cal L} | \CO ) \equiv | [H, \CO] )
\;.
\ee

 If we start with a `small' operator, containing few local degrees of freedom,  each nested commutator with the Hamiltonian tends to increase  its size. For a $k$-local Hamiltonian,  containing products of less than  $k$ local degrees of freedom, we expect the size of  ${\cal O}_n$ to be of order $n\,k$ for large values of $n$.  
 In  \cite{altman}, the authors build upon this remark and define complexity in terms of the `average' number of commutators which are required to construct a given operator $\CO$, starting with an initial operator ${\cal O}_0$. More precisely, if $\CO$  is in the vector space generated by the set of ${\cal O}_n$ operators seeded by ${\cal O}_0$ and $H$, we can write
 \be\label{gen}
 |{\cal O} ) = \sum_n i^n\, \varphi_n \,|{\cal O}_n )
 \ee
 for some complex coefficients $\varphi_n$. A measure of the typical number of $H$-commutators required to build $\CO$ is then 
 \be\label{kc}
 C_K ({\cal O}) = ( {\cal O} | \,n\,| {\cal O} ) = \sum_n n |\varphi_n|^2 \;,
 \ee
 and referred to as K-complexity, for its implicit dependence on the construction of the Krylov basis.  It was shown in \cite{altman} that this definition is  close to operator size for the SYK model
 in the thermodynamic limit. 
 
 Applying the general definition (\ref{kc}) to the time-evolved operator ${\cal O}_t = e^{itH}\,{\cal O}_0 \,e^{-itH}$ with initial condition ${\cal O}_0$, we are led to a natural notion of time-dependent K-complexity:
 \be\label{kct}
 C_K (t) = \sum_n \,n\,|\varphi_n (t) |^2\;,
 \ee
 which depends implicitly on the seed operator ${\cal O}_0$. The time-dependent components $\varphi_n (t)$ are obtained by solving the Heisenberg equation of motion, $\partial_t {\cal O}_t = i \,[H, {\cal O}_t]$   which, written in the Krylov basis,   takes the form 
  \be\label{recur}
 {\partial_t \varphi}_n = b_n\, \varphi_{n-1} - b_{n+1}\, \varphi_{n+1}
 \;,
 \ee
 with boundary condition $\varphi_{-1} (t) =0$. 
 
A given pattern of growth of Lanczos coefficients as a function of $n$ translates into a characteristic growth of complexity. For instance, it is shown  in \cite{altman}  that a system with an asymptotic large-$n$    law \footnote{In  local $(1+1)$-dimensional lattice systems, the linear law is modified by a logarithmic correction, namely  (\ref{chaos}) is replaced by  $b_n \sim \alpha n/\log n  + \dots$ (cf. \cite{altman}).}
  \be\label{chaos}
 b_n \approx \alpha \,n  \;,
 \ee
 accumulates K-complexity at an exponential rate:
 \be\label{expc}
 C_K (t) \sim e^{2\alpha t}\;.
 \ee
 A benchmark example of this behavior is the SYK model, for which $2\alpha = \lambda$ is the Lyapunov exponent revealed in OTOC correlations. It is then natural to propose (\ref{chaos}) as a criterion for local quantum chaos, since explicit evaluation of Lanczos coefficients in various  integrable systems yield softer asymptotic laws of the form
  \be\label{integ}
b_n \sim \alpha\, n^{\delta}\;, \qquad 0< \delta < 1\;.
\ee
In these cases, K-complexity has  a milder, powerlike growth: $C_K (t) \sim (\alpha t)^{1\over 1-\delta}$.

It  is useful to find relations between the patterns of growth of Lanczos coefficients and more familiar objects, such as  correlation functions. Let us consider the time autocorrelation 
\be\label{cor}
G(t) = ({\cal O}_0 \,| {\cal O}_t )  = {1\over {\cal N}}\; \Tr \left({\cal O}_0^\dagger \,{\cal O}_t \right)
\;,
\ee
which coincides with the standard Wightman correlation function at infinite temperature. In the thermodynamic limit, ${\cal N} \rightarrow \infty$, the Fourier transform 
\be\label{ft}
{\widetilde G} (\omega) = \int dt \,e^{-i\omega t} \,G(t) \;,
\ee
develops a non-trivial analytic structure. In particular, the singularities  closest to the real axis are located at $\pm i \pi/(2\alpha)$, where $\alpha$ is the slope coefficient  in (\ref{chaos}), and ${\widetilde G}(\omega)$ decays exponentially along the real axis with the law (cf. \cite{bookofruth}), 
\be\label{expd}
{\widetilde G}(\omega) \sim e^{-\pi |\omega|/2\alpha} \;.
\ee
More generally, a growth law of the form (\ref{integ}) translates into a decay ${\widetilde G}(\omega) \sim \exp(-|\omega/\omega_0|^{1/\delta})$, i.e. the sharper is the decay of the spectral function, the  milder is the  growth of the Lanczos coefficients. In the case  that the $b_n$ have a finite asymptotic limit $\lim_{n\to \infty} b_n = b_\infty$, it turns out that the spectral function has compact support in the finite interval $[-2b_\infty, 2b_\infty ]$. 

There is a direct relation between the Lanczos coefficients and the moments of the Liouvillian, 
\be\label{mo}
\mu_{n} = ( {\cal O}_0 \,| {\cal L}^{n} \,| {\cal O}_0 ) = \int{d\omega \over 2\pi} \;\omega^{n} \; {\widetilde G}(\omega)
\;,\ee
which in turn control the Taylor series of the autocorrelation function (only even moments contribute for Hermitian operators)
\be\label{moments}
G(t) = \sum_{n=0}^\infty \mu_{2n} {(it)^{2n} \over (2n)!}
\;.\ee
The relation between $b_n$ and $\mu_{2n}$ involves intricate  combinatorics, but there is a  lower bound 
\be\label{lowerb}
\mu_{2n} \geq b_1^2 \,b_2^2 \, \cdots b_n^2 
\;.
\ee
Furthermore, if the sequence of $b_n$ is non-decreasing, there is  also  an upper bound
\be\label{upperb}
\mu_{2n} \leq C_n\, b_1^2 \,b_2^2 \, \cdots b_n^2 
\;,
\ee
where $C_n = {(2n)! \over n! (n+1)!}$ is the $n$-th Catalan number. In particular, for a $b_n$   sequence  which is  non-decreasing  and asymptotic to $b_\infty$, one has 
\be\label{estim}
\mu_{2n} \sim (c\,b_\infty)^{2n +o(n)}\;, \qquad c>1\,,\qquad {\rm as} \;\; n\rightarrow \infty\;, 
\ee
where we have used the large-$n$ asymptotic form of the Catalan numbers $C_n \approx 4^n \,n^{-3/2}$. The notation $o(n)$ in the exponent stands for any terms with large $n$ growth slower than linear, such as fractional powers or logarithms.  

If the $b_n$ sequence is not strictly increasing, and yet $b_\infty$ exists,  then there is a non-decreasing sequence which approximates  $b_n$ asymptotically as $n\rightarrow \infty$. Hence,  we expect the estimate (\ref{estim}) to be qualitatively good provided the Lanczos sequence has a finite limit $b_\infty$.

\section{K-complexity of scramblers: fast and finite}

\noindent

In systems with a finite-dimensional Hilbert space,  K-complexity is necessarily bounded by the dimensionality of the operator space,
$C_K  \leq {\cal N}^{\,2}$.  Saturation of this bound is not guaranteed, as the Krylov basis may terminate its iterative construction  before it spans the whole operator space. Still,  for sufficiently generic choices of initial operator ${\cal O}_0$ and Hamiltonian $H$,   we expect that $n_{\rm max} $ does not lie far below ${\cal N}$.   To see this, consider the basis of operators 
\be\label{eles}
L_{(ab)} = \sqrt{\cal N} \,|\,E_a \rangle \langle E_b \,|\;,
\ee 
where $|E_a \rangle$ denotes the exact energy eigenstate with eigenvalue $E_a$. The ${\cal N}^{\,2}$ operators $L_{(ab)} $ define a basis of the operator space which is  orthonormal with respect to the inner product (\ref{inner}). The components of ${\cal O}_t$ in this 
  basis are proportional to   its matrix elements in the exact energy basis:
  $$
  (L_{(ab)} \,| {\cal O}_t ) ={1\over \sqrt{\cal N}} \, \langle \,E_a \,| {\cal O}_t \, | \,E_b \,\rangle\;,
  $$
  which at the same time can be written as 
\be\label{toro}
\left\langle E_a \,| \,{\cal O}_t \,| \,E_b \right\rangle = e^{i(E_a -E_b) t} \, \left\langle E_a \,| \,{\cal O}_0 \,|E_b \right\rangle\;.
\ee
For sufficiently generic initial operator, there are $O({\cal N}^{\,2})$ non-vanishing matrix elements, which remain non-vanishing at all times.  Thus, the 
  `supervector'  $|{\cal O}_t )$ has $O({\cal N}^{\,2})$  
 non-vanishing projections $(L_{(ab)} \,| {\cal O}_t ) $. Although the Krylov basis is rotated with respect to (\ref{eles}),  it is  natural to expect that the number of non-vanishing K-components $({\cal O}_n \,|\,{\cal O}_t )$ will also be of $O({\cal N}^{\,2})$. Furthermore,  for generic values of the energies $E_a$, the ${\cal N}$ independent phases $e^{-itE_a}$ describe an ergodic motion on a real ${\cal N}$-dimensional torus, which is embedded in the operator space by the equation (\ref{toro}).  Hence, $|{\cal O}_t )$ lies on an ${\cal N}$-dimensional submanifold and we can conclude that 
 $$n_{\rm max} = {\cal N}^{\,O(1)} = e^{O(S)} $$
 for systems with $S$ degrees of freedom, and generic choices of $H$ and ${\cal O}_0$.

 The computation of  K-complexities  requires the evaluation of  (\ref{kct}) once we know the amplitudes $\varphi_n (t)$. These in turn are obtained by solving (\ref{recur}). Therefore, it is the structure of the sequence $b_n$ what determines the relevant dynamical regimes in the growth of K-complexity. In a typical fast scrambler, such as the SYK model, small operators grow in size at an exponential rate $\exp(\lambda t)$, where $\lambda \approx 2\alpha$ is the Lyapunov exponent. In other words, for small operators,  operator size is roughly equivalent to K-complexity. 

We can regard the operator as `scrambled' when it has spread, in order of magnitude, over the whole system. For a fast scrambler with $S$ local degrees of freedom, this happens at the  familiar time scale  $t_* \sim \lambda^{-1} \,\log \,S$ \cite{sekinosuss}. 
The value of the K-complexity  at the scrambling time is of order 
\be\label{scrt}
C_K (t_*) \sim n_* \sim S \;.
\ee
For   systems with $O(S)$ lattice sites  and a finite-dimensional Hilbert space one  has ${\cal N} \sim e^{O(S)}$.  
Since $S \ll e^{O(S)} $,  it follows that  K-complexity has  an enormous scope for  growth beyond the `scrambling value'. Should the complexity continue to grow exponentially fast for $t>t_*$, it would saturate 
 in a time of order $S$.       In the next section we use the ETH hypothesis  to argue that this estimate is far from  correct.

\subsection{The ETH estimate}

\noindent

For systems which scramble less efficiently than a `fast' scrambler, one expects the scrambling time to scale like a power of $S$, rather than a logarithm, but the intuitive relation between K-complexity and operator size  suggests that the complexity at the scrambling time continues to satisfy (\ref{scrt}). Hence, the wide gap between the complexity at scrambling, $C_K (t_*) \sim S$, and the maximal complexity, of order $ e^{O(S)}$, should be a general feature of any system with finite degrees of freedom.

The rate of  K-complexity growth after scrambling depends on  the  form of the $b_n$ coefficients for $n\gg S$. These can be constrained from the behavior of the moments: 
\be\label{momeg}
\mu_{2n} =  \int{d\omega \over 2\pi} \;\omega^{2n} \; {\widetilde G}(\omega)
\;.\ee
From the spectral decomposition of the correlation function, 
\be\label{specgg}
{\widetilde G}(\omega) = {1\over {\cal N}} \sum_{a,b} \left|{\cal O}_{ab} \right|^2 \;2\pi\,\delta(\omega- (E_a - E_b))\;,
\ee
we obtain the expression:
\be\label{eeth}
\mu_{2n} = {1\over {\cal N}} \sum_{a,b} (E_a - E_b)^{2n} \,| {\cal O}_{ab} |^2\;,
\ee
where ${\cal O}_{ab} = \left\langle E_a | \,{\cal O}_0\,|E_b\right\rangle$ denote the matrix elements of the initial operator in the exact energy basis. These matrix elements can be used to characterize a degree of quantum chaos.  For  operators whose expectation values and correlations approach thermal values at long times, it is expected that    ${\cal O}_{ab}$  satisfy the Eigenstate Thermalization Hypothesis (ETH) \cite{peres, deutsch, srednicki}, which essentially says that the eigenbases of ${\cal O}_0$ and $H$ are uncorrelated, related by a random unitary on the ${\cal N}$-dimensional Hilbert space. From this assumption it follows that off-diagonal matrix elements contributing to (\ref{eeth}) 
 have the form
\be\label{eth}
{\cal O}_{ab} ={1\over \sqrt{\cal N}} \; F(E_a, E_b)   \;R_{ab} \;, \qquad a\neq b
\;,
\ee
where $R_{ab}$ is a random matrix whose entries have mean zero and unit variance. The form factor $F$ carries the information about the normalization of the operator and  is assumed to depend smoothly on the energies of the states. Plugging this ansatz into
the spectral expression (\ref{eeth}) we thus find
\be\label{etha}
\mu_{2n} \approx {1\over {\cal N}^{\,2}} \; \sum_{a, b} \, (E_a - E_b)^{2n} \;F(E_a, E_b)  
\;.\ee

For $n\gg S$ the energy sum tends to be dominated by the largest possible energy differences. For a system with $S$ degrees of freedom and extensive energy, the maximum energy difference 
is of order $\Lambda S$, where $\Lambda$ is the UV cutoff.    The sum over energy eigenvalues in (\ref{etha}) appears to be controlled by the   form factor's  bandwidth  $\Gamma$, which is an intensive energy scale, not scaling with $S$, and set by the local frequency cutoff $\Lambda$. For instance, assuming an exponential form factor $F(\omega) \sim e^{-\omega/\Gamma}$, we would estimate the sum over energy differences $\omega = E_a -E_b$ in (\ref{etha}) as proportional to 
$$
 \int_{0}^{\infty} d\omega  \,\omega^{2n} \,e^{-\omega/\Gamma} \sim (2n)!  \,\Gamma^{\;2n +1} \;,
$$
However, a saddle point analysis shows that, for  $n  \gg S$,  this integral gets its main contribution from $\omega_c \sim 2n\Gamma \gg \Lambda S$. Precisely for $n\gg S$, the saddle point sits outside the actual integration range, and the integral must be approximated ignoring the form factor. Ultimately, this is a consequence of the smoothness of the form factor as a function of $\omega$, and should hold generally for any operator satisfying the ETH ansatz. In particular,  quasinormal behavior in time correlations is associated to Lorentzian profiles of the form 
 $F(\omega) \sim (\omega^2 + \Gamma^2)^{-1}$, with an even milder  damping of large frequencies, so that  the previous argument applies as well.

We conclude that, as $n\gg S$,  the moment sum  is controlled by the average of $(E_a - E_b)^{2n}$
over the energy band, i.e.
\be\label{mome}
\mu_{2n} \sim {1\over n^2} ( \Lambda\,S)^{2n} \;, \qquad {\rm as} \;\;n\gg S \;.
\ee
Going back to (\ref{estim}), this results in  Lanczos coefficients approaching  an asymptotic `plateau'  at height 
\be\label{as}
b_\infty \sim \Lambda S \;. 
\ee
For a fast scrambler,  $b_{n_*} \sim \alpha n_* \sim \alpha S$. Therefore, if all couplings are of order unity, the Lyapunov exponent must be itself of the order of the local characteristic frequency, $\lambda \sim 2\alpha \sim \Lambda$, within factors of order unity, and we are led to a very  simple picture for the Lanczos sequence: linear growth with slope $\lambda$ for $0<n<n_* $ morphing into an approximate plateau extending all the way to $n_{\rm max} \sim e^{O(S)}$.

The qualitative description of a fast scrambler, as determined by a Lyapunov exponent $\lambda \sim \Lambda$ and $S$ degrees of freedom, can be adapted more general situations where only a subset of degrees of freedom are `activated' in the scrambling process. This  occurs when considering a system at a finite temperature below the UV cutoff, $T<\Lambda$. In this case, on states of entropy $S$, the system can be described as having about one degree of freedom per thermal cell of size $\beta = T^{-1}$ participating in the scrambling process, with the rest of degrees of freedom effectively `frozen' in their ground state, and thus not contributing to the entropy. On such states of entropy $S$
and effective temperature $T$, the UV cutoff is effectively replaced by $T$, setting the scale of the Lyapunov exponent $\lambda \sim T$. 

\begin{figure}[t]
$$\includegraphics[width=\textwidth]{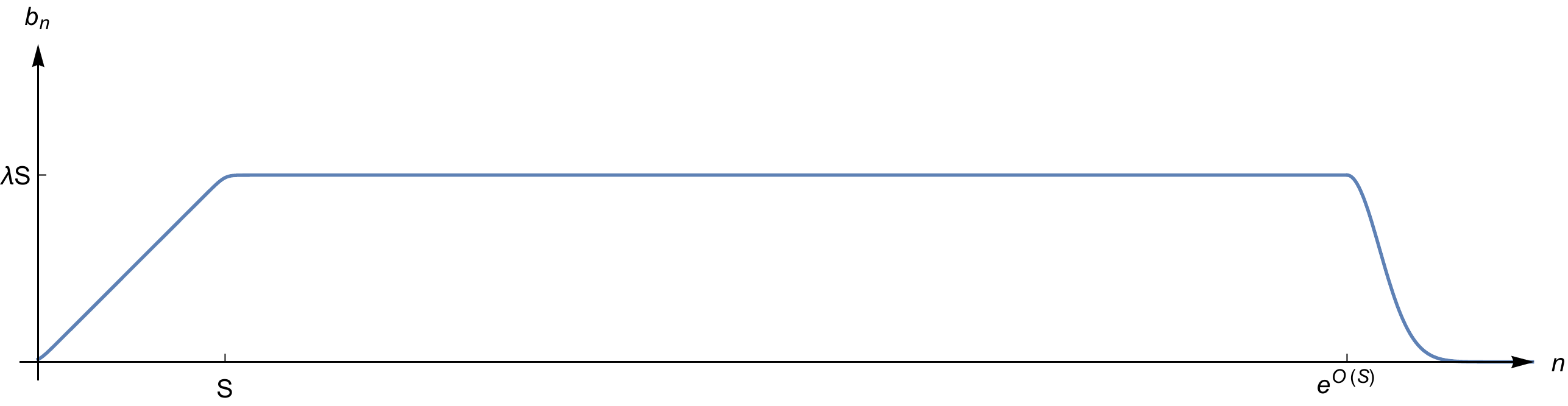} $$   
\begin{center}
\caption{\emph{Qualitative form of the Lanczos sequence for a fast scrambler with $S$ degrees of freedom and  Lyapunov exponent $\lambda$. It shows the linear growth characteristic of a fast scrambler, the long constant regime and a sharp turnoff at saturation. }\label{fig:lanczos}}
 \end{center}
\end{figure}

The qualitative band-structure of $b_n$ coefficients for a fast scrambler is  shown in  Fig. \ref{fig:lanczos}.  More generally, for systems with less efficient scrambling, the initial linear growth might be substituted by  (\ref{integ}), whereas the `post-scrambling' plateau for $n\gg S$ is expected to be rather general.   It would be very interesting to test the generality of this `Lanczos plateau' in numerical simulations of various models, such as SYK.

\subsection{Dynamics of K-complexity}

\noindent

The evolution of K-complexity for a fast scrambler with linear growth (\ref{chaos})  was studied in \cite{altman}. An analytic solution for the amplitudes $\varphi_n (t)$ exists for a formal choice of Lanczos coefficients given by 
 $b_n = \sqrt{n(n-1+\eta)} $. To simplify matters, we look at the exactly linear case,  corresponding to $\eta=1$, for which the solution reads 
\be\label{linsol}
\varphi_n (t) =  \tanh^n (\alpha t)\; {\rm sech} (\alpha t)\;.
\ee
An initially sharp peak at $n=0$ moves to higher $n$ exponentially fast: $n_{\rm peak} (t) \sim e^{2\alpha t}$. The overall height of the function at large $t$ is of order $e^{-\alpha t}$. Hence, the scrambling is very efficient at accessing `large' operators but at the same time it is also very efficient in randomizing the operator in the Krylov basis, leading to an essentially flat  $\varphi_n$  distribution with support on $[0, n_{\rm peak} (t)] $ and height of order $1/\sqrt{n_{\rm peak}(t)}$. 

The growth of complexity is largely controlled by the ballistic motion in $n$-space of the solution's  `wave front'. On the other hand, operator randomization depends on whether   a significant tail is left behind the wave front. For a discussion of the ballistic aspect, as well as the detailed matching between the pre-scrambling and post-scrambling regimes, it is useful to start with a continuum approximation.

Taking a coarse-grained look at the discrete function $\varphi_n (t)$, let us introduce a lattice cutoff $\varepsilon$ and a coordinate $x=\varepsilon \,n$, and define the interpolating functions $\varphi(x,t) = \varphi_n (t)$, $v(x) = 2\varepsilon \,b(\varepsilon n) = 2\varepsilon \,b_n$. A continuum form of the recursion relation (\ref{recur}) can be written as:
\be\label{contir}
\partial_t \varphi(x,t) = {1\over 2\varepsilon} \left[v(x) \, \varphi (x-\varepsilon) - v(x+\varepsilon) \,\varphi(x+\varepsilon) \right] \;.
\ee

Expanding now in powers of $\varepsilon$, we  find to leading order 
\be\label{conta}
\partial_t \varphi = -v(x)  \partial_x \varphi - {1\over 2}  \partial_x v(x) \,\varphi + O(\varepsilon )\;, 
\ee
a chiral wave equation with position-dependent velocity $v(x)$ and mass $\partial_x v(x) /2$. We can solve it by introducing a new coordinate $y$ by the relation $v(x) \partial_x = \partial_y$, and a rescaled amplitude
\be\label{resc}
\psi(y, t) = \sqrt{v(y)} \,\varphi(y, t)
\;,
\ee
which simplifies the chiral wave equation:
\be\label{wec}
(\partial_t + \partial_y ) \,\psi(y,t) = 0 +\dots\;,
\ee
 the dots standing for the neglected terms of higher order in the $\varepsilon$ expansion. The general solution of this equation
is given by
\be\label{ges}
\psi(y,t) = \psi_i (y-t)\;,
\ee
where $\psi_i (y) = \psi(y,0)$ is the initial condition. 

The rescaling (\ref{resc}) is also useful from the point of view of the intuition about probability distributions. From the discrete normalization condition $\sum_{n\geq 0} |\varphi_n |^2 =1$ we can derive the continuum analogs 
\be\label{contnorm}
1= {1\over \varepsilon} \int dx\, |\varphi(x)|^2 = {1\over \varepsilon} \int dy \,v(y) \,|\varphi(y)|^2 = {1\over \varepsilon} \int dy \,|\psi(y) |^2\;,
\ee
so that $\psi(y)$ is a naive probability amplitude in $y$ space, just as $\varphi(x)$ is a naive probability amplitude in $x$ space.

 The physics of (\ref{ges})  is that of a simple ballistic motion of the initial $\psi$-distribution towards positive values of $y$ at a constant velocity. The problem is solved once we know the change of variables between the $x$-frame and the $y$-frame.  The K-complexity as a function of time is given by
\be\label{kt}
C_K (t) = \sum_n n\,|\varphi_n (t)|^2 \approx {1\over \varepsilon^2} \int dx \,x\, |\varphi(x, t)|^2 = {1\over \varepsilon^2} \int dy\,x(y)\,|\psi(y, t)|^2 \;.
\ee
Using the general solution (\ref{ges}) in the last expression and changing variables $y\rightarrow y+t$ we find
\be\label{ktt}
C_K (t) = {1\over \varepsilon^2} \int dy \,x(y+t)\,|\psi_i (y)|^2\;. 
\ee
 There are various interesting cases to consider. 
 
 A fast scrambler with linear Lanczos growth has $v(x) = \lambda \,x$, where $\lambda =2\alpha$ is the Lyapunov exponent. The corresponding change of variables is
 \be\label{changes}
y = {1\over \lambda} \,\log (x/\varepsilon)\;,\qquad x= \varepsilon\; e^{\lambda\, y}
\ee
where we have chosen the additive normalization in $y$ for convenience. Notice that, in this case, the $y$ variable runs over the whole real line, whereas the $x$ variable is restricted to be positive. The scrambling solution for the $\varphi$ amplitude then reads
\be\label{scsol}
\varphi(x, t)_{\rm scrambling} = e^{-\lambda t /2} \varphi_i \left(x\,e^{-\lambda t}\right)\;.
\ee
An initial peak at $y=0$ for $\psi_i (y)$ will move ballistically as $y_p (t) = t$,  corresponding to an $x$-frame trajectory
\be\label{peakm}
x_p (t) = \varepsilon \,e^{\lambda\,t}\;,
\ee
which also controls the exponential growth of K-complexity. 

If the velocity has a logarithmic correction, as proposed in \cite{altman} for $(1+1)$-dimensional systems, 
$$
v(x) = {\lambda \,x \over \log (x/\varepsilon)}\;, 
$$
the corresponding frame map is
$$
x= \varepsilon\,e^{\sqrt{2\lambda y}}\;.
$$
and the distribution peak and K-complexity grow at a rate of order $\exp(\sqrt{2\lambda t})$. 

For systems with a less efficient scrambling, governed by (\ref{integ}) with $\delta <1$, the drift velocity is given by 
$$
v(x) = 2\alpha \,\left({x \over \varepsilon}\right)^\delta\;,
$$
leading to a change of variables
$$
y = {1\over 2\alpha} {1\over 1-\delta} \left( \left({x\over \varepsilon}\right)^{1-\delta} - 1 \right)\;, \qquad x= \varepsilon \left( 1+ 2(1-\delta) \alpha \,y\right)^{1\over 1-\delta}\;.
$$
and a  power-like complexity growth
proportional to $
 \left(\alpha \,t\right)^{1\over 1-\delta}
$. 

It is interesting to compare estimates of scrambling times based on the growth of K-complexity with other heuristic models of scrambling. If we define the scrambling time by the requirement that complexity reaches the size of the system, $C_K (t_*) \sim S$, then we have 
\be\label{sctm}
t_* \sim {1\over \alpha} \,S^{1-\delta} 
\;.
\ee
On the other hand, in $d$-spatial dimensions, ballistic scrambling takes a time of order $t_* \sim L$ for a system of size $L$. If we write
$S\sim (\alpha\,L)^d$ for the effective number of degrees of freedom (entropy) and $\alpha^{-1}$ for the effective dynamical time step, we have $t_* \sim \alpha^{-1} \,S^{1/d}$ for ballistic scrambling. If we model the scrambling by a diffusion process, characterized by a random walk of step $\alpha^{-1}$, we obtain instead $t_* \sim \alpha^{-1}\, S^{2/d}$. Then, we find the interesting correspondences
\be\label{corre}
\delta_{\rm ballistic} = 1-{1\over d}\;, \qquad \delta_{\rm diffusion} = 1-{2\over d}\;.
\ee

\subsubsection*{The post-scrambling regime}

\noindent

In the post-scrambling regime the $x$ and $y$ frames are simply proportional: $x(y) = v\,y$, where $v(x)$ is now approximately constant $v(x) \approx v = v_* $, and the amplitude $\varphi(x,t)$ just moves ballistically towards large $x$ with velocity $v_*$,
\be\label{afersc}
\varphi(x, t)_{\rm post-scrambling} = \varphi(x-v_* (t-t_*), t_*) \,,
\ee
the K-complexity also growing linearly. 

To summarize, in the simplest case of an SYK-like fast scrambler, with Lyapunov exponent $\lambda$ and $S$ extensive degrees of freedom, we have $v(x) \approx \lambda \,x$ in the scrambling band, $0<x< \varepsilon \,S$ and $v(x) = v_* \sim \lambda \,\varepsilon \,S$ in the post-scrambling band. 
In the scrambling period, $x(y) \approx \varepsilon \,e^{\lambda y} $ resulting in 
\be\label{kttps}
C_K (t)_{\rm scrambling} \approx e^{\lambda t} \,C_K (0) = e^{\lambda(t-t_*)} C_K (t_*)\;,
\ee
the expected exponential growth. If the initial operator is `small', the initial complexity is also small,  $C_K (0) = { O}(1)$ and $C_K (t_*) = O(S)$. On the other hand, in the post-scrambling regime $x(y) \approx y\,v$ with constant $v=v_* = \varepsilon \lambda n_* \sim \varepsilon \lambda \,S$. At long times:
\be\label{kttpos}
C_K (t)_{\rm post-scrambling} \approx \lambda \,n_* \,(t-t_*) + C_K (t_*)  \sim \lambda \,S\, (t-t_*) + { O}(S)\;,
\ee
using the normalization condition (\ref{contnorm}). We conclude that the complexity grows exponentially fast during scrambling and only linearly after scrambling, with a rate of order $\lambda S$. The time scale for the amplitude to reach 
$n_{\rm max} $ is of order
\be\label{ktime}
t_K \sim \varepsilon \;{n_{\rm max} \over v} \sim {1\over \lambda S} \,e^{O(S)} \;. 
\ee

\begin{figure}[t]
$$\includegraphics[width=\textwidth]{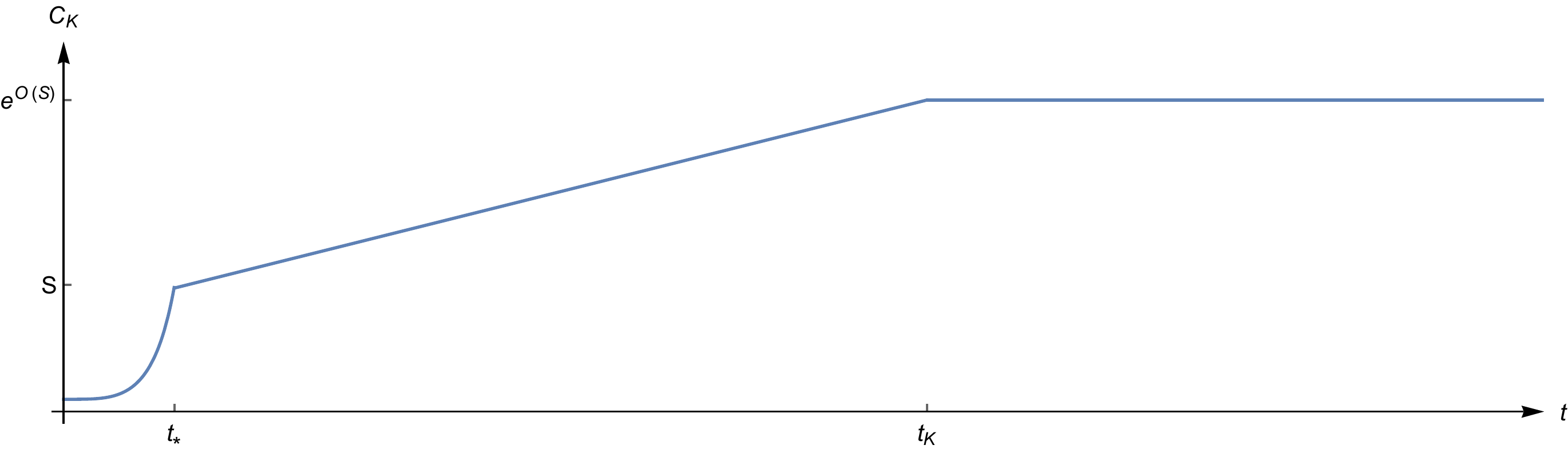} $$   
\begin{center}
\caption{\emph{Evolution of K-complexity for a fast scrambler of size $S$, featuring an exponential law in the pre-scrambling era $t<t_* \sim \lambda^{-1} \,\log\,S$, followed by a a linear law  in the post-scrambling era, up  to $t_K \sim e^{O(S)} / \lambda S$, when the complexity finally saturates. }\label{fig:ck}}
 \end{center}
\end{figure}

At times larger than $t_K$, the function $\psi(y, t)$ remains stuck near the endpoint, because the drift towards large values of $x$ will prevent the distribution from bouncing back.  This implies  that the complexity  eventually levels off and remains constant. Over extremely long time scales, however, we know that the solution of the  discrete equation (\ref{recur}) will necessarily  undergo Poincar\'e recurrences. The time scale for this to happen is of order 
\be\label{poinc}
t_P \sim {1\over \lambda S} \exp\left(e^{O(S)} \log (1/\epsilon)\right) \;,
\ee
where $\epsilon$ determines the precision with which we demand recurrence. We summarize the qualitative behavior of the K-complexity for a fast scrambler in Fig. \ref{fig:ck}.

\section{Operator randomization and K-entropy}
\noindent
 
 Having established the existence of a very long post-scrambling era of linear K-complexity growth, we now begin a more detailed study of this dynamical regime. In particular, we discuss the degree of randomization of the operator ${\cal O}_t$, when expanded in the Krylov basis. For this purpose, we shall introduce the notion of K-entropy. In order to motivate its definition, we momentarily go back to the scrambling period. 
 
 The exact solution  (\ref{linsol}) describes two a priori independent phenomena: there is an exponentially fast growth of K-complexity and at the same time there is an efficient randomization of the operator over the time-dependent span of the
 Krylov operator set. This is intuitively clear from the qualitative form of (\ref{linsol}), which eventually looks like a uniform distribution of size $n_{\rm peak}$ and amplitude $1/\sqrt{n_{\rm peak}}$. A more formal characterization of this uniformity is given by the `operator entropy' or K-entropy, which we define by 
\be\label{opentropy}
S_K = -\sum_{n\geq 0} |\varphi_n |^2 \, \log \,|\varphi_n |^2\;.
\ee
If the $\varphi_n$ amplitude is very peaked at a particular value of $n$, large or small, the K-entropy is small. On the other hand, if the distribution is completely uniform over the interval $[0, n_M]$, then $S_K = \log \,(n_M)$. Applying the definition (\ref{opentropy}) to
(\ref{linsol}) we can determine the growth of K-entropy to be expected from a typical fast scrambler. The result of a  numerical evaluation is a linear growth with    slope close to $2\alpha = \lambda$. Hence, the scrambling dynamics increases K-complexity at an exponential rate, and also increases K-entropy at a linear rate. 

It turns out that the linear growth of K-entropy for a fast scrambler is   captured by the  continuum solution of the leading equation (\ref{conta}).  
The  continuum versions of K-entropy in both $x$-frame and $y$-frame  are 
\be\label{opentropyc}
S_K = -{1\over \varepsilon} \int dx \,|\varphi(x,t) |^2 \,\log \, |\varphi (x,t) |^2 = -{1\over \varepsilon} \int dy\; |\psi(y,t)|^2 \, \log \left({|\psi(y,t)|^2 \over v(y)}\right) \;.
\ee
Extracting the velocity-dependent term in the $y$-frame expression, we have
\be\label{openplus}
S_K = -{1\over \varepsilon} \int dy \,|\psi(y,t)|^2 \, \log |\psi(y,t)|^2
+ {1\over \varepsilon} \int dy\; \log \,v(y) \,|\psi(y)|^2 \;.
\ee

In the leading continuum approximation, any $y$-frame solution has the form $\psi(y,t) = \psi_i (y-t)$. Hence, the first term is time-independent, whereas the second term computes the average of $\log( v(y))$ over the operator probability distribution. In periods where the complexity growth is accelerated, such as the scrambling period of a fast scrambler, there is entropy production. Inserting the leading continuous  solution (\ref{scsol}) of the scrambling regime into  (\ref{opentropyc})  one obtains 
\be\label{scentp}
S_K (t)_{\rm scrambling} = S_K (0) + \lambda \, t\;, 
\ee
which matches   the numerical evaluation for the exact solution (\ref{linsol}).
 This means that the simple chiral wave equation with a mass term (\ref{conta}) actually gives a very accurate description of the scrambling regime, not only accounting for the growth of K-complexity, but also capturing quantitatively the growth of K-entropy.

In the post-scrambling period where $v(x) \approx$ constant, the mass term in (\ref{conta}) is negligible and the amplitude propagates ballistically in both frames.  Therefore, the  leading order term in the continuum approximation
to the amplitude does not detect any significant growth of the K- entropy. 
We now turn to analyse what can seen at some higher orders.

\subsection{The continuum amplitude at post-scrambling}
\noindent

We have seen that, while  operator randomization is well  accounted for in the continuum approximation for the scrambling regime, it is completely missed at leading order in the post-scrambling regime. It is an important question to determine whether K-entropy can be produced at all during the enormously long post-scrambling era.   

 In this section we show that the next-to-leading approximation to the evolution equation (\ref{contir}) already begins to incorporate the randomization effect, but ultimately falls short of the goal. 
Carrying the short distance expansion of (\ref{contir}) to higher orders one finds (\ref{conta}), with further corrections on the right hand side. At order $\varepsilon$ there is  a term
$$
-{1\over 2} \,\varepsilon \,\partial_x v(x) \,\partial_x \varphi(x,t) \;,
$$
which corrects the velocity to $v_{\rm eff} (x) = v(x) - \varepsilon \, \partial_x v(x) /2 $. This is a small effect in the scrambling regime and completely negligible in the post-scrambling regime. At order $\varepsilon^2$ we find two terms:
$$
-{1\over 4}\, \varepsilon^2 \,\partial_x v(x) \,\partial_x^2 \varphi (x, t) -{1\over 6} \,\varepsilon^2 \,v(x)\,\partial_x^3 \varphi(x,t)\,.
$$
The first term is a diffusion contribution with the wrong sign of the diffusion constant, and it only acts for a small time in the scrambling era. The second term is active throughout the long post-scrambling era and thus corresponds to the leading correction which is in principle capable of incorporating a broadening effect. 

\begin{figure}[t]
$$\includegraphics[width=3in]{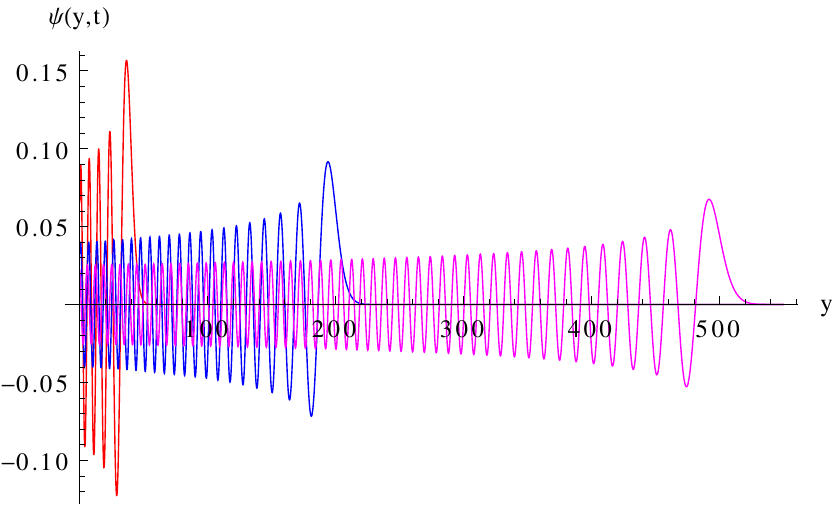} $$   
\begin{center}
\caption{\emph{Plot of the Airy function (\ref{appt}) for $t=40$ (red), $t=200$ (blue) and $t=500$ (magenta). Notice the very efficient randomization. }\label{fig:airy}}
 \end{center}
\end{figure}

Let us then consider the $O(\varepsilon^2)$-corrected equation in the post-scrambling regime $t>t_*$ and in the $y$-frame, 
\be\label{broadeq}
(\partial_t + \partial_y ) \psi(y,t) = -\gamma \,\partial_y^3 \,\psi(y,t)\;,
\ee
 written in terms of the rescaled amplitude $\psi(y, t) = \sqrt{v(y)} \varphi(y, t)$, which has  standard $L^2$ norm in the $y$-frame. The coefficient controlling the new term is certainly small:
\be\label{gammaco}
\gamma = {1\over 6} {\varepsilon^2 \over v^2} \sim {1 \over (\lambda S)^2} \;,
\ee
where we have used that  $v=\varepsilon\, \lambda \,n_* \sim \varepsilon \,\lambda \, S$ in the post-scrambling regime. 
In order to solve (\ref{broadeq}) we seek a solution of Fourier form
\be\label{fou}
\psi(y, t) = \int {dk \over 2\pi} \,\psi_k \, e^{-i\omega_k (t-t_*) + iky} \;,
\ee
with dispersion $\omega_k = k - \gamma\, k^3$. Let us set an initial condition at $t=t_*$, specifying  the amplitude as  $\psi(y, t_*) = \psi_i (y)$, which is just the Fourier transform of $\psi_k$. The solution reads  
\be\label{gensoll}
\psi(y,t) = \int dy' \, \psi_i (y') \int {dk \over 2\pi} \;e^{ik\,(y-y'-\Delta t) +i\gamma \,\Delta t \,k^3} \;,
\ee
where $\Delta t = t-t_*$. By the rescaling $k\rightarrow k / (3\gamma \Delta t)^{1/3}$ we can evaluate  the momentum integral in terms of the Airy function to obtain
\be\label{airyform}
\psi(y, t) = \int dy' \,\psi_i \left[(3\gamma \Delta t)^{1/3 }\,y'\,\right] \; {\rm Ai}  (z-y') \;,
\ee
where 
$$
z= {y - \Delta t \over (3\gamma \Delta t )^{1/3}} \;.
$$
It is already clear from this expression that this approximation is beginning to capture the randomization effect, due to the properties of the Airy function. To see this, let us  consider an initial delta-function pulse, 
$$
\psi_i (y) =A\, \delta(y-y_*)\;, 
$$
leading to a solution 
\be\label{appt}
\psi(y, t) = {A\over (3\gamma \Delta t)^{1/3} }\, {\rm Ai} \left[{\Delta y - \Delta t \over (3\gamma \Delta t)^{1/3}} \right]\;,
\ee
where  $\Delta y = y-y_*$. The constant $A$ is fixed  by requiring the correct normalization of $\psi(y,t)$. Evaluating the asymptotics long after the ballistic front $y \sim t$ has passed, i.e.  $\Delta t \gg \Delta y$, one obtains 
\be\label{appp}
\psi(y, t) \approx {A \over \sqrt{\pi}}  {1\over (3\gamma \Delta t)^{1/4}} {1 \over (\Delta t - \Delta y)^{1/4}} \,\sin \left[ {2 \over 3} \left({\Delta t -\Delta y \over (3\gamma \Delta t)^{1/3}} \right)^{3/2} + {\pi \over 4}\right]\;, 
\ee
for $0< \Delta y < \Delta t$,  and essentially zero otherwise. 
The normalization condition
$$
1= {1\over \varepsilon} \int dy \, |\psi(y)|^2 
$$
fixes the order of magnitude of  the constant to be
$
A \sim (3\gamma)^{1/4} \sqrt{\varepsilon} 
$, so that the operator amplitude looks like a rapidly oscillating function over the interval $0< \Delta y < \Delta t$ of the form 
\be\label{airytail}
\psi_{\rm tail} \sim \sqrt{\varepsilon \over \Delta t} \;\times {\rm Osc}_{\;[0,t]}\;,
\ee
where ${\rm Osc}_{\;[0,t]}$ stands for the oscillation component with unit amplitude (a cosine function) and support on the interval $[0, t]$. Converting back to the $x$-frame amplitude $\varphi = \psi /\sqrt{v}$ we have an oscillating function
\be\label{aairytail}
\varphi_{\rm tail}\sim \sqrt{\varepsilon \over v\Delta t} \;\times {\rm Osc}_{\;[0,vt]} \;,
\ee
with amplitude of order $\sqrt{ \varepsilon/vt}$ and 
and support on the  interval  $[0,vt]$. 
 This result is interesting, since it shows perfectly efficient randomization (cf. Fig. \ref{fig:airy}). The very flat and long tail yields a K-entropy of order
 \be\label{kena}
 S_K \sim  \log (vt/\varepsilon) =\log ( 2bt)
 \ee
  at long times. However, a delta-function initial condition is not a realistic starting point for the post-scrambling regime. First, such a singular initial configuration is beyond the regime of applicability of the low-derivative approximations to  (\ref{contir}). Second, it was argued that a period of fast-scrambling with $S$ degrees of freedom outputs a distribution with an $x$-width of order $x_* \sim \varepsilon \,n_* \sim \varepsilon S \gg \varepsilon$. Hence, in order to check if the present approximation captures randomization, we must input an initial distribution of width $\delta x \sim \varepsilon\,S$. Equivalently, in the $y$-frame at post-scrambling this amounts to $\delta y \sim \delta x / v_* \sim \lambda^{-1}$. 

Picking a gaussian ansatz for the normalized $y$-frame distribution, 
\be\label{gausans}
\psi_i (y, t_*) = \pi^{-1/4} \sqrt{\varepsilon \over \delta}  \,\exp\left(-{(y-y_*)^2 \over 2\delta^2}\right) \;, 
\ee
 the integral (\ref{airyform}) may be evaluated exactly to obtain\footnote{To compute this integral, it is useful to go back to (\ref{gensoll}) and do the $y'$ integral first. In the resulting momentum integral, an appropriate shift the integration variable $k$ produces  an integral representation of the Airy function.}
\be\label{exactg}
\psi(y, t) = {\pi^{1/4} \sqrt{2\varepsilon \delta} \over (3\gamma \Delta t)^{1/3}} \; e^B \;{\rm Ai} \left[{\Delta y - \Delta t +C \over (3\gamma \Delta t)^{1/3}} \right]\;,
\ee
where 
\be\label{constab}
B=-{\delta^2 \over 6\gamma} \left(1-{\Delta y \over \Delta t}\right) + {\delta^6 \over 108 \gamma^2 \Delta t^2} \;, \qquad C= {\delta^4 \over 12 \gamma \Delta t}\;.
\ee
Looking at the long-time tail we focus on the region of large $\Delta y $ with $\Delta t \gg \Delta y$, so that only the first term in $B$ remains relevant as a correction to the Airy function profile. This term induces a suppression of order $\exp(-\delta^2 / 6\gamma)$ on the tail amplitude. Putting all factors together one finally finds 
\be\label{newtail}
\varphi_{\rm tail} \sim \sqrt{\delta\, \lambda \, n_*} \;e^{-(\delta\, \lambda \,n_*)^2 } \;{1 \over \sqrt{2b  t}} \; \times \,{\rm Osc}_{\;[0,2bt]}\;,
\ee
up to $O(1)$ factors. 

\begin{figure}[t]
$$\includegraphics[width=3in]{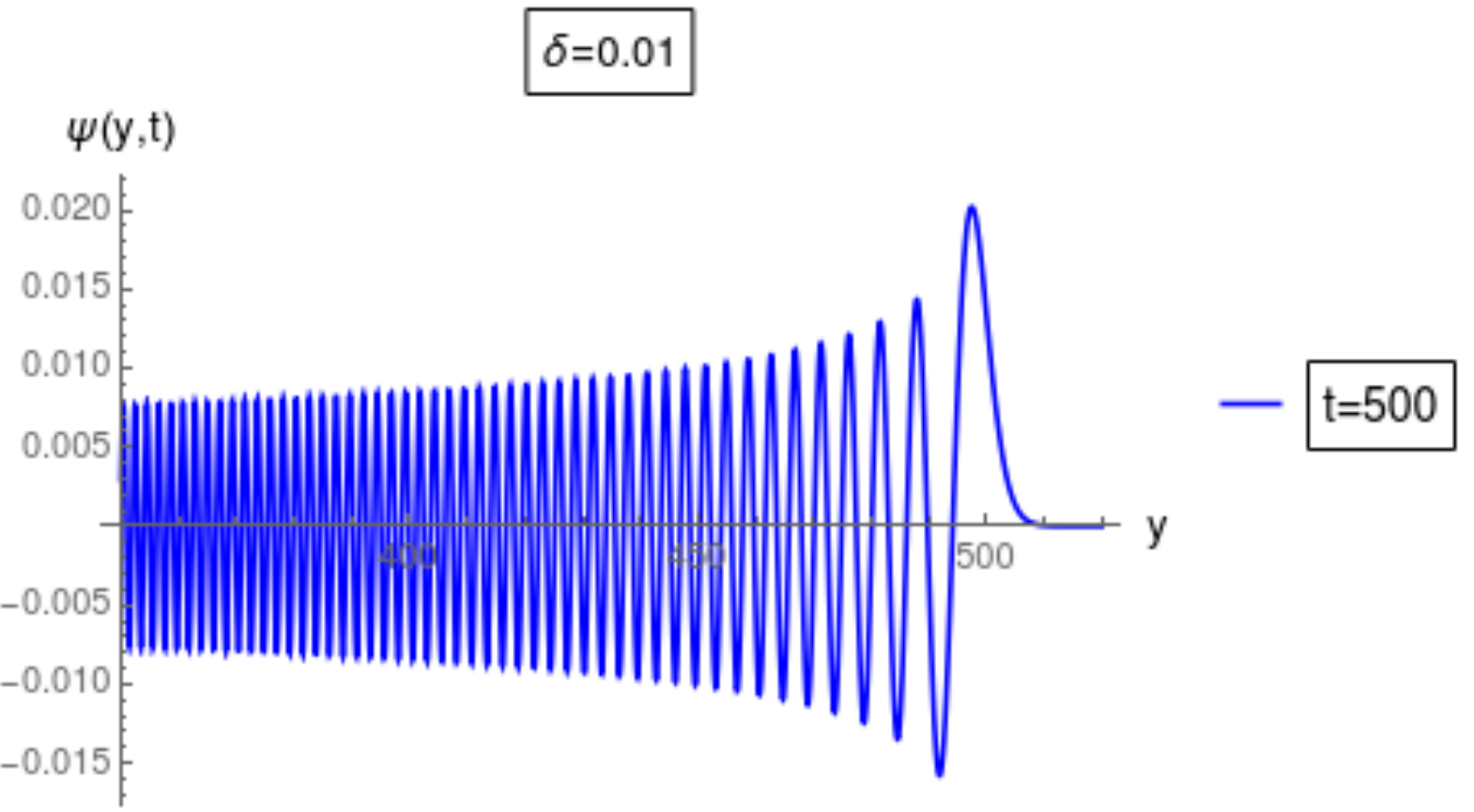} $$   
\begin{center}
\caption{\emph{ The amplitude (\ref{exactg}) for a very narrow initial pulse, $\delta = 10^{-2}$ in units of the Lyapunov exponent, is very similar to the amplitude for an initial delta-function pulse.}\label{fig:gaussianairy}}
 \end{center}
\end{figure} 

\begin{figure}[t]
$$\includegraphics[width=3in]{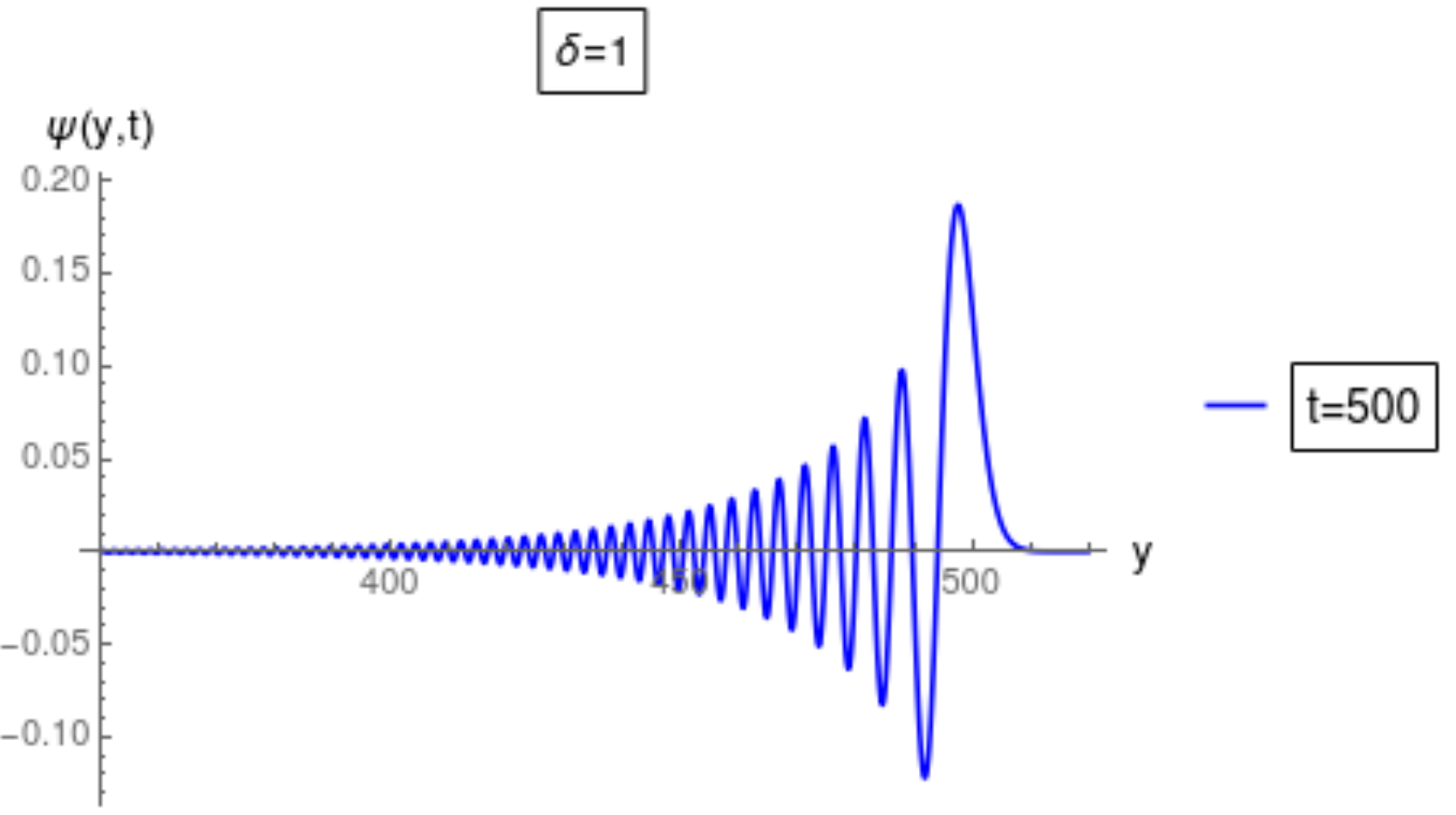} $$   
\begin{center}
\caption{\emph{The amplitude (\ref{exactg}) for a wide initial pulse, with $\delta =1$ in units of the Lyapunov exponent, exhibits an exponentially damped tail.  }\label{fig:gaussiandamped}}
 \end{center}
\end{figure}

We conclude that, unless we pick a lattice-size distribution, with $\delta \sim 1/\lambda S$, the randomization is all but washed out when looking at smooth signals. In particular, for the choice of width $\delta\sim 1/\lambda$,  which corresponds to an initial scrambling period of time $t_* = \lambda^{-1}\;\log\,S$,  the tail is exponentially suppressed, 
\be\label{lastt}
\varphi_{\rm tail} \sim \sqrt{S} \,e^{-S^2} \;{1\over \sqrt{2bt}} \;\times \,{\rm Osc}_{\;[0,vt]}\;,
\ee
and the propagation is essentially ballistic. We show the difference between the two choices of initial width in Figs. \ref{fig:gaussianairy} and \ref{fig:gaussiandamped}. 

On the other hand, the fact that randomization arises when the signal is extrapolated to cutoff scales,  beyond the domain where we trust the equation (\ref{broadeq}), suggests that perhaps randomization is a true property of the discrete evolution equation.

\subsection{The discrete amplitude at post-scrambling}

\noindent

In search for K-entropy production in the post-scrambling regime, we return to the discrete problem (\ref{recur}), which becomes 
\be\label{consb}
{\partial_t \varphi}_n = b(\varphi_{n-1} - \varphi_{n+1})\;, 
\ee
when  the Lanczos coefficients are approximated  by a constant $b_n \approx b$.  In the physical situation of interest, this equation holds for $n>n_* \sim S$, and the solution must be matched to a solution of the scrambling regime, such as (\ref{linsol}). 

\begin{figure}[t]
$$\includegraphics[width=3in]{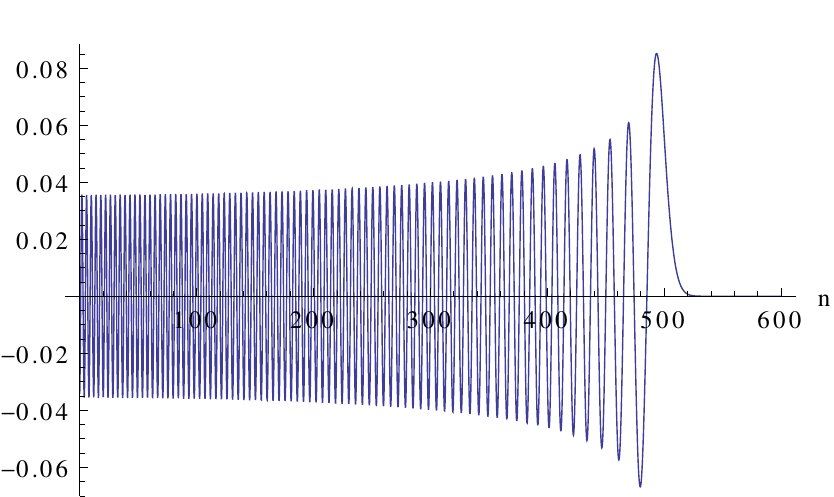} $$   
\begin{center}
\caption{\emph{Plot of the Bessel function $J_n (2bt)$ as a function of $n>0$ for $2bt = 500$. }\label{fig:Besselfunn}}
 \end{center}
\end{figure} 

Ignoring boundary conditions for the time being, a particular solution of (\ref{consb}) is just a Bessel function:
\be\label{bessel}
\varphi_n (t) =  J_n (2bt)\;.
\ee
 It has the correct normalization at $t=0$, with all amplitudes vanishing except $\varphi_0 (0) =1$. Therefore the Bessel functions describe the spread of a distribution which begins sharply localized at the origin. A glance at the plot in Fig. \ref{fig:Besselfunn}  reveals that randomization is very efficient, featuring a tail similar to that of the Airy function found in the last section. Using the so-called `approximation by tangents' (cf. \cite{GR}) we can write, for $n$ large at
fixed ratio $2bt / n >1$: 
\be\label{apprb}
J_n (2bt) \approx {1\over (4b^2 t^2 -n^2)^{1/4}}\;\cos\left[ \sqrt{4b^2 t^2 -n^2} - n \,a- {\pi \over 4} \right] \;,
\ee
where $a = {\rm arc\;tan} \sqrt{4b^2 t^2 -n^2}$. As the distribution moves to large $n$ at constant velocity, equal to $2b$, there is a rapidly oscillating tail with almost flat envelope and height  of order $(4b^2 t^2 -n^2)^{1/4} $. Therefore, the Bessel function restricted to positive $n$ behaves qualitatively as the Airy function, featuring an oscillating tail with amplitude of order  $1/\sqrt{2bt}$, supported on the interval  $[0,2bt]$.

The Bessel function amplitude has 
however in this case unphysical features,  because it leaks into the negative $n$ axis, as the ansatz (\ref{bessel}) fails to satisfy the correct boundary condition $\varphi_{-1} =0$. This implies that the probability density $|\varphi_n |^2 $ is not conserved on the physical configurations with $n\geq 0$. The problem can be fixed by a  superposition of two Bessel functions: 
\be\label{ruthf}
R_n (2bt) = J_n (2bt) + J_{n+2} (2bt)\;,
\ee
which vanishes identically at $n=-1$ for all times, as one can   verify using the identity $J_{-n} (z) = (-1)^n \,J_n (z) $. As a result, $R_{-1} (2bt)=0$ is effectively a `Dirichlet' condition separating the dynamics of the physical region $n\geq 0$ and the dynamics of the unphysical region $n<-1$. Furthermore, 
$R_n (t=0) = \delta_{n,0} + \delta_{n, -2}$ and,  since one can now consistently restrict attention to positive values of $n$, it follows  that (\ref{ruthf})  does satisfy the physical conditions of being narrowly localized at $t=0$ and permanently confined in the $n\geq 0$ region. 

\begin{figure}[t]
$$\includegraphics[width=3in]{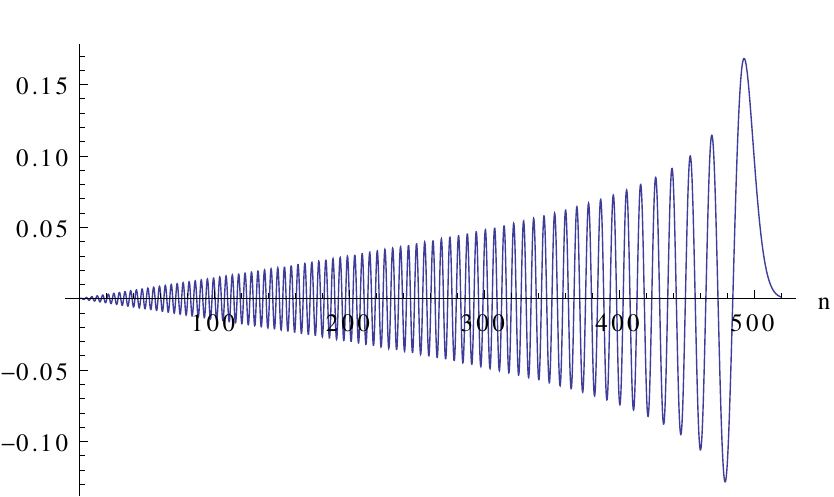} $$   
\begin{center}
\caption{\emph{Plot of  $R_n (2bt)$ as a function of $n>0$ for $2bt = 500$. }\label{fig:ruthf}}
 \end{center}
\end{figure}

The function (\ref{ruthf}) can be rewritten as 
\be\label{rutho}
R_n (2bt) = {n+1 \over bt} \,J_{n+1} (2bt)\;,
\ee
a from that makes manifest a linear enveloping behavior at large $n$, as shown in Fig. \ref{fig:ruthf}.  Despite this accumulation of probability at the higher end of the $n$ spectrum, one can check using the form (\ref{rutho}) that the K-entropy {\it does} grow at a logarithmic rate 
$
S_K [R_n (2bt)] \propto \log \,(2bt) 
$ at long times, the hallmark of a good operator randomization. 

The function $R_n (2bt)$ locates  the initial pulse right next to the Dirichlet condition. In order to better simulate  the type of configuration prepared by a previous scrambling period, it is  convenient to engineer analogs of the $R_n$ function with initial pulses located at any desired position. These `displaced' pulses can be manufactured by generalizing  (\ref{ruthf}) into 
\be\label{ruthk}
R^{(k)}_n (2bt) = J_{n-k} (2bt) + (-1)^k\,J_{n+k+2} (2bt) 
\ee
for any non-negative integer $k$. 
These functions meet the goal since they vanish at $n=-1$ for all times and  $R^{(k)}_n (0) = \delta_{n, k} + (-1)^k \,\delta_{n, -k-2}$. Hence, we have a function which starts with a unit pulse at any $n=k\geq 0$,  while remaining confined to the $n\geq 0$ domain at all times, the original pulse function (\ref{ruthf}) corresponding to the particular case of $k=0$.

\begin{figure}[t]
$$\includegraphics[width=3in]{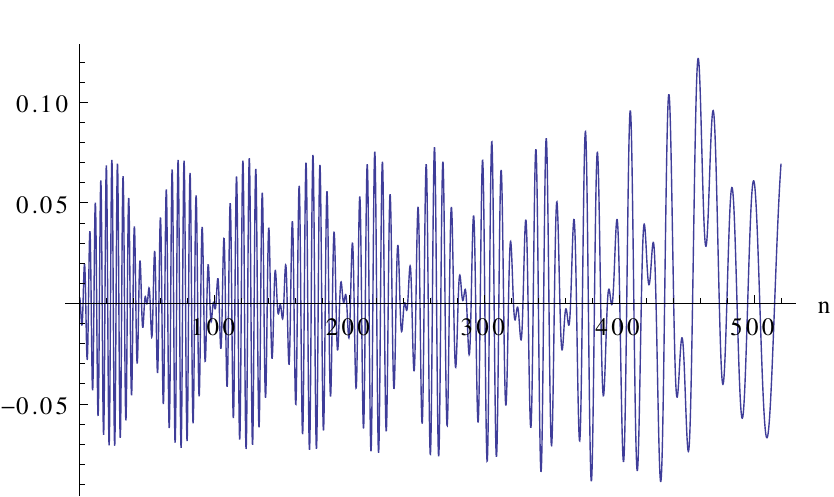} $$   
\begin{center}
\caption{\emph{Plot of  $R^{(k)}_n (2bt)$ as a function of $n>0$ for $k=30$ and $2bt = 500$. }\label{fig:ruthk}}
 \end{center}
\end{figure}

For generic values of $k$ and long times, the $k$-pulse functions $R^{(k)}_n (2bt)$  look  like modulated Bessel functions, i.e. they display a tail of average height of order $1/\sqrt{2bt}$ and are supported on the ballistic domain bounded by $n_t \sim 2bt$ (cf. Fig. \ref{fig:ruthk}), therefore, they also feature logarithmically increasing K-entropies. 

\begin{figure}[t]
$$\includegraphics[width=3in]{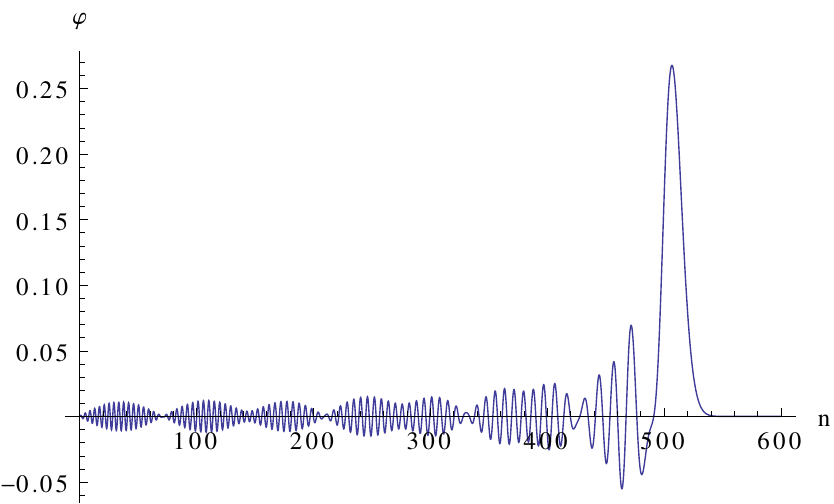} $$   
\begin{center}
\caption{\emph{Plot of the amplitude (\ref{wideruth}) as a function of $n>0$ for $2bt=500$ and  an initial square pulse of width $K_0 =20$.}\label{fig:wideruth}}
 \end{center}
\end{figure}

With these ingredients in place, we are ready to discuss the more realistic case of an initial pulse with arbitrary width $K_0$. This can be achieved by a superposition of $k$-pulses
\be\label{wideruth}
\varphi_n (t) = \sum_{k=0}^{K_0 -1} \alpha_k \,R^{(k)}_n (2bt) \;, \qquad \sum_{k=0}^{K_0 -1} |\alpha_k |^2 =1\;.
\ee
In particular, choosing $K_0 \sim S$ simulates the kind of signal that is prepared by a previous period of fast scrambling. To simplify matters, let us consider a square pulse with $\alpha_k = 1/\sqrt{K_0}$. 

\begin{figure}[t]
$$\includegraphics[width=3in]{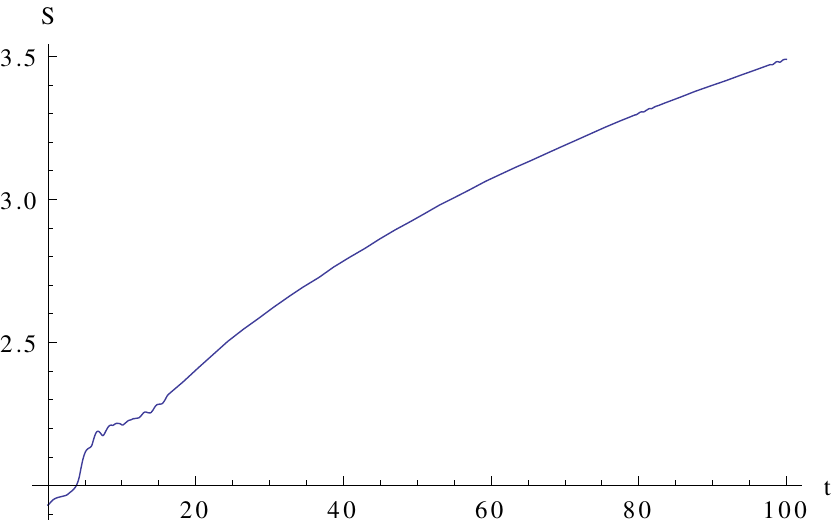} $$   
\begin{center}
\caption{\emph{Growth of K-entropy for an initial square pulse of width $K_0 = 5$. Notice the asymptotic logarithmic growth and
the initial finite-size effects due to the details of the square pulse.}\label{fig:logplot}}
 \end{center}
\end{figure} 

An example of the long time evolution of such a pulse is shown in Fig. \ref{fig:wideruth}. We observe a stable peak which propagates ballistically and an approximately uniform tail obtained by averaging over tails  of single-pulse functions. Assuming that the phases of each single-pulse function add up randomly, we estimate that a randomization tail exists with height of order $1/\sqrt{2bt}$ and width of order $2bt$, leading to a logarithmic growth of  K-entropy: $S_K (t) \sim \log (2bt)$. This logarithmic growth  for the
K-entropy can be confirmed  by direct numerical evaluation (cf. Fig. \ref{fig:logplot}).

\begin{figure}[t]
$$\includegraphics[width=5in]{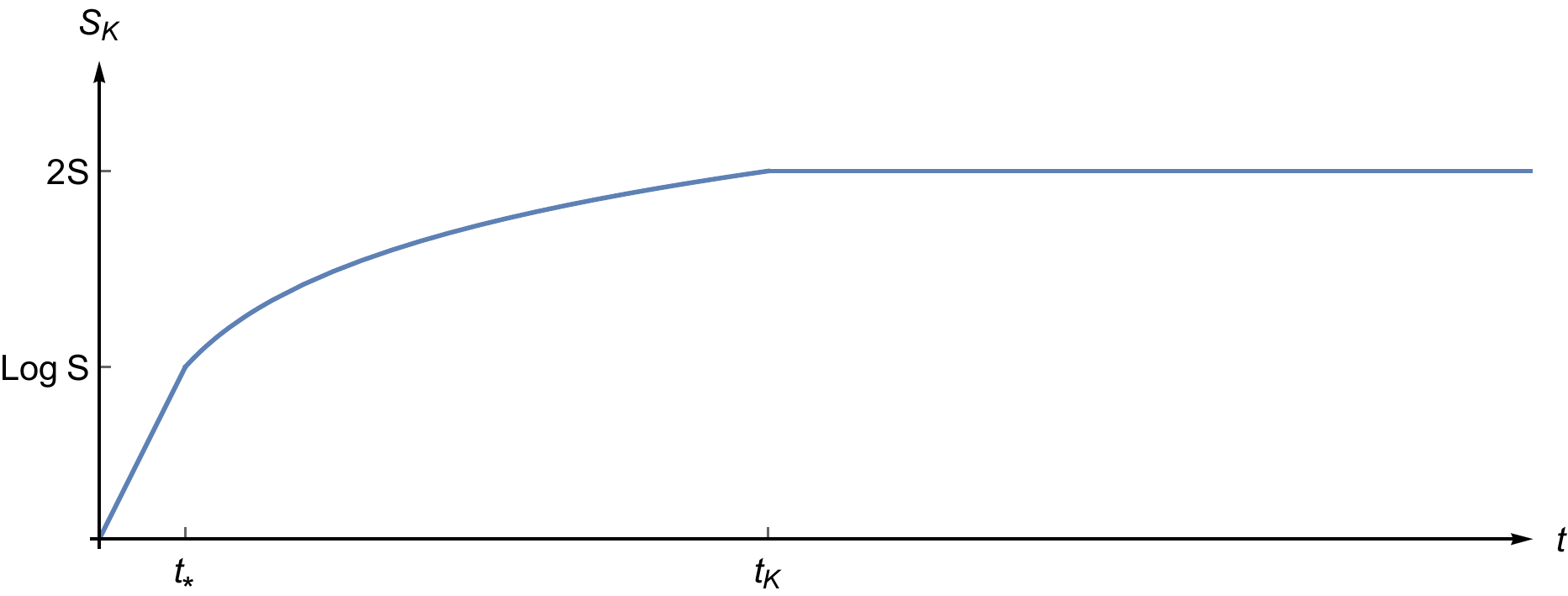} $$   
\begin{center}
\caption{\emph{Sketch of the K-entropy dynamics in a fast scrambler with $S$ degrees of freedom and Lyapunov exponent $\lambda$. A linear growth proportional to $\lambda\,t$ during scrambling is followed by a logarithmic increase in the post-scrambling era, according to a scaling $\log(2S\lambda t)$, and a final saturation beyond times of order $t_K$.}\label{fig:ken}}
 \end{center}
\end{figure}

The conclusion is that randomization does occur in order of magnitude. There is a persistent ballistic component which makes an $O(1)$ fraction of the normalization, but the K-entropy at long times is dominated by the oscillating tail. Eventually, after times of order $t_K$, the K-entropy becomes of order $\log (n_{\rm max})$, thereby growing  from $O(\log S)$ at $t_*$ to $O(S)$ by the exponential time scale $t_K$. A qualitative picture of the K-entropy dynamics in a fast scrambler is presented in Fig. \ref{fig:ken}.

\section{Conclusions}

\noindent

In this paper we have explored the long-time behavior of K-complexity, an algebraic notion of operator complexity which relies on an effective dimensionality of a linear subspace containing the operator's time evolution. This concept was introduced in \cite{altman} as 
a useful characterization of chaotic behavior, in the sense of being governed by the same Lyapunov exponent as OTOC correlators. 

Using the  Eigenstate Thermalization Hypothesis as  a starting point, we have argued that K-complexity grows linearly at late times, after the system has been scrambled, with a rate which is extensive in the size $S$ of the system. Eventually, the K-complexity must saturate at a maximum bounded by $e^{O(S)}$, in a time also proportional to $e^{O(S)}$,   and stays approximately constant  thereafter, until Poincar\'e recurrences begin to show up at times scaling as a double exponential of the entropy, $\exp(e^{O(S)})$. 

We furthermore notice that, during  the exponentially long post-scrambling period when K-complexity grows linearly, the operator is randomized in order of magnitude. This can be  characterized by the logarithmic growth of the K-entropy, which measures the degree of uniformity of the amplitudes $\varphi_n (t)$.
More precisely, we find numerical evidence for a growth law of the form
$$
S_K \sim \log\,(2bt) \;,
$$
as $bt\gg 1$, 
where $b$ denotes  the asymptotic value of the Lanczos sequence. At complexity saturation, the K-entropy also saturates at a value 
of order $\log (n_{\rm max})  = O(S)$. It would be interesting to study the consequences of this randomization on the long time behavior of correlation functions, along the lines of \cite{br, ss, yoshida, ssb}.

The outstanding open question regarding these results is the holographic representation of K-complexity. During the scrambling period, there is an approximate correspondence between K-complexity and operator size. There are proposals for  concrete relations between operator size and  bulk quantities \cite{susn}  \cite{javi}. In  these examples, the holographic map is specified between the process of particle free-fall towards a horizon and a scrambling process in the holographic dual. The natural expectation is that a period of linear growth of complexity should be associated to properties of the motion in the interior of the black hole.

\section{Acknowledgments}
\label{ackn}
J.L.F. Barbon would like to thank the Hebrew University in Jerusalem and  IHES for hospitality during the preparation of this work. R. Shir would like to thank the IFT-Madrid for hospitality during the preparation of this work.
 R Sinha would like to thank the Strings 2019 conference in Brussels, Belgium where the work was first presented.
 
 The work of J.L.F. Barbon and R. Sinha is partially supported by the Spanish Research Agency (Agencia Estatal de Investigaci\'on) through the grants IFT Centro de Excelencia Severo Ochoa SEV-2016-0597,  FPA2015-65480-P and 
PGC2018-095976-B-C21. 
The work of E. Rabinovici and R. Shir is partially supported by the Israeli Science Foundation Center of
Excellence and by the I Core Program  ``The Quantum Universe," sponsored by the
25
Planning and Budgeting Committee and the Israeli Science Foundation.


\end{document}